\documentclass[preprintnumbers,eqsecnum,aps,prd,epsf,showpacs,nofootinbib,twocolumn]{revtex4}
\usepackage{latexsym,amsmath,amssymb}
\usepackage[dvips]{graphicx,color}
\usepackage{dcolumn,bm}    
\usepackage{subfigure}  

\def\til#1{\tilde{#1}} 
\begin{document}
\thispagestyle{empty}
\title{Full nonlinear growing and decaying modes 
of superhorizon curvature perturbations}

\preprint{RESCEU-26-10}
\author{Yu-ichi Takamizu}
\email{takamizu_at_resceu.s.u-tokyo.ac.jp}

\author{Jun'ichi Yokoyama}
\email{yokoyama_at_resceu.s.u-tokyo.ac.jp}

\affiliation{
\\
Research Center for the Early Universe (RESCEU), Graduate
School of Science, The University of Tokyo, Tokyo 113-0033, Japan }
\date{\today}

\begin{abstract}
 We clarify the behavior of curvature perturbations in 
a nonlinear theory in case the inflaton temporarily 
stops during inflation. We focus on the evolution 
of curvature perturbation on superhorizon scales by adopting 
the spatial gradient expansion and show that 
the nonlinear theory, called the 
{\it beyond} $\delta N$-formalism for a general single scalar field 
as the next-leading order in the expansion. 
Both the leading-order in the 
expansion ($\delta N$-formalism) and our nonlinear theory 
include the solutions of full-nonlinear 
orders in the standard perturbative expansion. 
Additionally, in our formalism, we can deal with the 
time evolution in contrast to $\delta N$-formalism, where 
curvature perturbations remain just constant, and show 
decaying modes do not couple 
with growing modes as similar to the case with linear theory. 
We can conclude that although the decaying mode diverges when 
$\dot{\phi}$ vanishes, there appears no trouble for 
both the linear and nonlinear theory since these modes will 
vanish at late times.
\end{abstract}
\maketitle

\section{Introduction}
Recent observations of the cosmic microwave background anisotropy \cite{Komatsu:2010fb} show very good agreement of the observational data with the 
prediction of standard inflationary cosmology, that is, adiabatic 
Gaussian random primordial fluctuations with an almost scale-invariant 
spectrum generated from quantum fluctuations of an inflaton field during inflation \cite{Guth:1980zm,Starobinsky:1980te,Sato:1980yn,Mukhanov:1981xt,Hawking:1982cz,Starobinsky:1982ee,Guth:1982ec}. 
The amplitude of curvature perturbation on a comoving 
slicing, ${\cal R}_c$, is given by the formula ${\cal R}_c\approx 
H^2/|\dot{\phi}|$ evaluated at time of horizon crossing $t=t_k$ when the wavenumber $k$ satisfying $k=aH$, 
where $\phi$ and $H$ are the inflaton and the Hubble parameter during inflation, respectively. The reason why it gives an almost scale-invariant spectrum is that 
both $H$ and $\dot{\phi}$ change very slowly during slow-roll inflaton. 

The main purpose of this paper is to clarify what happens when the inflaton stops during that, namely, 
$\dot{\phi}\approx 0$, using a nonlinear perturbation theory. 
Such a situation naturally occurs in oscillating inflation or the 
chaotic new inflation models 
\cite{Damour:1997cb, Liddle:1998pz,Yokoyama:1998pt}. 
For example, it has been shown by Damour and Mukhanov that 
oscillating inflation is realized as the inflaton oscillates around a minimum of a nonconvex potential \cite{Damour:1997cb}. In another example of chaotic new inflation \cite{Yokoyama:1998pt},  it has been pointed that 
the inflaton changes its direction of motion if model parameters are 
approximately chosen. 

In both examples, if we use the above formula for the amplitude of 
primordial curvature perturbation, it apparently diverges when 
$\dot{\phi}$ vanishes. However, Seto, Yokoyama and Kodama \cite{Seto:1999jc} have shown that even in the case that 
slow-roll conditions are violated, 
if a new formula is applied to this case, 
the amplitude has still finite value, where the time derivative of the 
scalar field is replaced by the potential gradient given as 
${\cal R}_c\propto 3H^3/V'(\phi)$ at $t=t_k$. In this study, 
they have investigated the evolution of curvature perturbations in the linear theory and shown a decaying mode can diverge at and around temporary stopping of the inflaton \cite{Seto:1999jc, Kodama:1996jh}. 
However, since the decaying mode vanishes 
at sufficient late times, there appears no trouble in the linear perturbation theory. 

When 
we take nonlinear effects into account, the decaying modes can couple with 
the growing modes in general to 
convert into growing modes through such effects. 
Therefore the ill-behavior of the decaying mode may leave an observable 
trace if nonlinear perturbation is incorporated. 
The purpose of this paper is to clarify the behavior of curvature perturbations in a nonlinear theory in case 
the inflaton temporarily stops during inflation. 

In order to incorporate nonlinearity of curvature perturbation, 
we focus on the evolution on superhorizon scales and 
consider a nonlinear cosmological perturbation theory by adopting 
a gradient expansion approach \cite{Salopek:1990jq}. 
As for the leading-order in the expansion, $\delta N$-formalism \cite{Starobinsky:1986fxa, Nambu:1994hu, Sasaki:1995aw,Lyth:2005fi} is a 
powerful tool to calculate the nonlinearity 
of primordial curvature perturbations 
(recently much attention to as their non-Gaussianity \cite{Bartolo:2004if}) 
since it includes 
the solutions of full-nonlinear orders in the standard perturbative expansion, 
but this is just a 
lowest-order and in this formalism, we should ignore all decaying modes. 
Therefore, we have to use the next-leading order in the expansion, 
which was recently formulated by one of us, the so-called 
{\it beyond} $\delta N$-formalism \cite{Takamizu:2010xy}.  
In our formalism, which we will briefly review in the following section, 
there exists decaying and growing modes, having their time-dependences and they lead to time variations of superhorizon curvature perturbations. 
We will show such decaying mode also diverges, but they will vanish due to inflationary expansion in the same way 
as in the linear theory, when $\dot{\phi}$ vanishes. 

The rest of the paper is organized as follows. In Sec. II, we review the 
full-nonlinear cosmological perturbation theory of superhorizon curvature perturbations. Then we discuss the growing and decaying modes in both the linear and nonlinear theories in Sec. III and discuss what happens on temporary stopping 
of the inflaton in Sec. IV. Section V is devoted to the conclusion.

\section{Beyond $\delta N$-Formalism }
\label{}
In this section, 
we will briefly review the nonlinear theory of 
cosmological perturbations valid up to $O(\epsilon^2)$ in the spatial
gradient expansion and follow the previous works \cite{Takamizu:2010xy,Takamizu:2008ra}, where 
$\epsilon$ is the ratio of the Hubble length scale $1/H$ to the characteristic length scale of perturbations $L$, used as a 
small expansion parameter, $\epsilon\equiv 1/(HL)$, of the 
superhorizon scales. 
First of all, we show the main result in our formula for the 
nonlinear curvature perturbation, ${\cal R}_c^{\rm NL}$, 
\begin{eqnarray}
{{\cal R}_c^{\rm NL}}''+2 {z'\over z} 
{{\cal R}_c^{\rm NL}}' +{c_s^2\over 4} K^{(2)}[\,
{\cal R}_c^{\rm NL}\,]=O(\epsilon^4)\,,
\label{eq: basic eq for NL}
\end{eqnarray}
which shows two full-nonlinear effects; 
\begin{enumerate}
\item Nonlinear variable: ${\cal R}_c^{\rm NL}$ including full-nonlinear curvature perturbation, $\delta N$
\item Source term: $K^{(2)}[{\cal R}_c^{\rm NL}]$ is a nonlinear function of curvature perturbations. 
\end{enumerate}
In (\ref{eq: basic eq for NL}), the prime denotes conformal time derivative and $z$ is a well-known Mukhanov-Sasaki variable which will be seen later as (\ref{def: variable-z}).  
The explicit forms of both 
the definition of ${\cal R}_c^{\rm NL}$ and the source term 
$K^{(2)}[X]$, that is 
the Ricci scalar of the metric $X$,  
will be also seen later, in (\ref{def0: nonlinear variable zeta}) 
and in (\ref{def: K2}), respectively.
Of course, 
in the linear limit, it can be reduced to the well-known equation
for the curvature perturbation on comoving hypersurfaces \cite{Mukhanov:1990me},
\begin{eqnarray}
{{\cal R}^{\rm Lin}_c}''+2{z'\over z} {{\cal R}^{\rm Lin}_c}'
-c_s^2\,\Delta[{\cal R}^{\rm Lin}_c]=0\,.
\label{eq: linear eq R}
\end{eqnarray}

We will briefly summarize our formula and 
show the above results in the following. 
Throughout this paper we consider a minimally-coupled single scalar field
described by an action of the form
%
\begin{equation}
 I = \int d^4x\sqrt{-g}P(X,\phi), 
\end{equation}
where $X=-g^{\mu\nu}\partial_{\mu}\phi\partial_{\nu}\phi$. 
Note that we do not 
assume the explicit forms of both kinetic term and its 
potential, that can be given as arbitrary function of $P(X,\phi)$. 

We adopt the ADM decomposition and employ the gradient expansion. 
In the ADM decomposition, the metric is expressed as 
\begin{equation}
 ds^2 = - \alpha^2 dt^2 + \gamma_{ij}(dx^i+\beta^idt)(dx^j+\beta^jdt), 
\end{equation}
where $\alpha$ is the lapse function, $\beta^i$ is the shift vector and
Latin indices run over $1, 2, 3$. 
The equations of motion corresponding to $\alpha$ and $\beta^i$ 
lead to constraint equations. Components of
the spatial metric $\gamma_{ij}$ are dynamical variables and the corresponding equations of motion  are reduced to 
a set of first-order differential equations with
respect to the time $t$. We introduce the extrinsic
curvature $K_{ij}$ defined by  
%
\begin{equation}
 K_{ij} =
  -\frac{1}{2\alpha}\left(\partial_t\gamma_{ij}-D_i\beta_j-D_j\beta_i\right),
  \label{eqn:def-K}
\end{equation}
where $D$ is the covariant derivative compatible with the spatial metric
$\gamma_{ij}$. As a result, 
the basic equations are reduced to 
the first-order equations for the dynamical variables $(\gamma_{ij}$,$K_{ij})$,  with the two constraint equations (the so-called 
Hamiltonian and Momentum constraint). 
We further 
decompose the spatial metric and the extrinsic curvature as 
%
\begin{eqnarray}
 \gamma_{ij} & = & a^2e^{2 \zeta}\tilde{\gamma}_{ij}, 
  \nonumber\\
 K_{ij} & = & 
  a^2e^{2 \zeta}\left(\frac{1}{3}K\tilde{\gamma}_{ij}
	    +\tilde{A}_{ij}\right),
	     \label{eqn:decompose-metric}
\end{eqnarray}
where $a(t)$ is the scale factor of the background FRW universe and 
${\rm det} \tilde{\gamma}_{ij}=1$. 

Next, we will employ the gradient expansion. In this approach 
we introduce a flat FRW universe 
($a(t)$, $\phi_0(t)$) as a background. As discussed in the first part of 
this section, we consider the perturbations on superhorizon scales, that 
is, $L$ is longer than the Hubble length scale $1/H$ of the background, 
i.e.  $HL\gg 1$. 
Therefore,
we consider $\epsilon\equiv 1/(HL)$ as a small expansion 
parameter and systematically expand our equations by $\epsilon$, considering a spatial 
derivative acted on perturbations is of order $O(\epsilon)$. 

We assume the condition for the gradient expansion: 
\begin{equation}
 \partial_t\tilde{\gamma}_{ij} = O(\epsilon^2). 
 \label{eqn:assumption-gamma}
\end{equation}
This corresponds to assuming the absence of any decaying modes
at the leading-order in the expansion, namely, the absence
of spatially homogeneous anisotropy.
This is justified in most of the inflationary models in which
the number of $e$-folds of inflation $N$ is much larger than
the number required to solve the horizon and flatness problem,
$N\gg 60$. 
This assumption is 
sufficient to allow us discuss behavior of decaying modes when the 
inflaton stops, since all time dependent solutions at 
the leading order are reduced to just decaying modes and there exists 
no observable trace at late times.

When we focus on a contribution arising from the scalar-type 
perturbations, we may choose the gauge in which $\tilde{\gamma}_{ij}$ approaches the flat metric, 
\begin{eqnarray}
\tilde{\gamma}_{ij} \,(t\to \infty) =\delta_{ij},
\label{eq: gamma-ij t-infty}
\end{eqnarray}
where in reality the limit $t\to \infty$ may be reasonably interpreted as an 
epoch close to the end of inflation. 
We take the {\it comoving slicing, time-orthogonal} gauge: 
\begin{eqnarray}
\delta \phi_c(t,x^i)=\beta_c^i(t,x^i)=O(\epsilon^3), 
\label{def: comoving-time-orth}
\end{eqnarray}
where $\delta \phi\equiv\phi-\phi_0$ denotes a fluctuation of a scalar field. 
The subscript $c$ denotes this gauge throughout this paper.

Now we turn to the problem of properly defining a nonlinear
curvature perturbation to $O(\epsilon^2)$ accuracy. 
Hereafter we will use the expression ${\cal R}_c$ on 
comoving slices to denote it. 
Let us consider 
the linear curvature perturbation which is  
given as 
\begin{eqnarray}
{\cal R}^{\rm Lin}=\left(H^{\rm Lin}_L+{H^{\rm Lin}_T\over 3}\right)Y,
\label{def: linear curvature}
\end{eqnarray}
where, following the notation in \cite{Kodama:1985bj},
the spatial metric in the linear limit is expressed as 
\begin{eqnarray}
\gamma_{ij}=a^2(\delta_{ij}
+2 H^{\rm Lin}_LY \delta_{ij}+2 H^{\rm Lin}_T Y_{ij})\,,
\end{eqnarray}
with $Y$ being scalar harmonics with eigenvalue $k^2$ in Fourier space 
satisfying
\begin{eqnarray}
(\Delta + k^2)Y=0\,,
\end{eqnarray}
and 
\begin{eqnarray}
Y_{ij}=k^{-2} \left[
\partial_i \partial_j -{1\over 3} \delta_{ij} \Delta \right] Y\,.
\label{eq: scalar harmonics}
\end{eqnarray}
These expressions in the linear theory 
correspond to the metric components in our
notation as
\begin{eqnarray}
{\zeta}=
 H^{\rm Lin}_L Y,~~\til{\gamma}_{ij} = \delta_{ij} + 2H^{\rm Lin}_T Y_{ij}.
\label{def: notition linear quantity}
\end{eqnarray}
Notice that the variable ${\zeta}_c$ reduces to 
${\cal R}^{\rm Lin}_c$ at leading-order in the gradient expansion, 
but not at second-order as (\ref{def: linear curvature}) 
and it will be also similar to the nonlinear theory. 

Thus to define a nonlinear generalization of the linear curvature
perturbation~(\ref{def: linear curvature}), we need nonlinear generalizations
of $H_LY$ and $H_TY$. Our nonlinear ${\zeta}$ is an
apparent natural generalization of $H^{\rm Lin}_LY$,
\begin{eqnarray}
H_LY=\zeta\,.
\end{eqnarray}
As for $H_TY$, however, the generalization is non-trivial.
It corresponds to the $O(\epsilon^2)$ part of $\tilde\gamma_{ij}$ and 
we have obtained a general solution of the dynamical equation for $\tilde{\gamma}_{ij}$ as a first-order differential equation in \cite{Takamizu:2010xy,Takamizu:2008ra} and 
the time-dependent part includes the following solution;
\begin{eqnarray}
\tilde{\gamma}_{ij}(t) \ni C^{(2)}_{ij}\int \frac{dt'}{a^3(t')},
\label{sol: time-dep-C}
\end{eqnarray}
with the Momentum constraint:
\begin{eqnarray}
e^{3\ell^{(0)}}\partial_iC^{(2)} & = & 
  6 f_{(0)}^{jk}\partial_j
  \left[e^{3\ell^{(0)}}C^{(2)}_{ki}\right].
  \label{sol: time-dep-C2}
\end{eqnarray}
The explicit forms of solutions can be seen in \cite{Takamizu:2008ra}. 
Here we attach the superscript $(m)$ to a quantity of $O(\epsilon^m)$, and 
both $\ell^{(0)}$ and $f^{(0)}_{ij}$ will be denoted 
as the leading-order metric in (\ref{eqn:psi-leading}) and 
(\ref{eqn:gammatilde-leading}). Our aim is to 
derive the scalar-type solution $C^{(2)}$ from 
the tensor $C^{(2)}_{ij}$ in (\ref{sol: time-dep-C}) by using 
(\ref{sol: time-dep-C2}). As shown in \cite{Takamizu:2010xy}, it can be 
done by introducing
the inverse Laplacian operator $\Delta^{-1}$ on the flat background and 
we defined the nonlinear generalization of $H_TY$ as
\begin{eqnarray}
H_TY=E\equiv
-\frac{3}{4}\Delta^{-1}\left[\partial^i e^{-3 \ell^{(0)}}
\partial^j e^{3 \ell^{(0)}}(\ln \tilde{\gamma})_{ij} \right].
\label{def: nonlinear HT0}
\end{eqnarray}
It is easy to see that $E\ni C^{(2)}$ which we expected.

At leading-order, the only non-trivial quantities for the spatial 
metric, $\zeta$ and $\tilde{\gamma}_{ij}$, are given by 
%
\begin{equation}
 \zeta = \ell^{(0)}(x^k) + O(\epsilon^2),
  \label{eqn:psi-leading}
\end{equation}
and 
\begin{equation}
 \tilde{\gamma}_{ij} = f^{(0)}_{ij}(x^k) + O(\epsilon^2),
  \label{eqn:gammatilde-leading}
\end{equation}
where $\ell^{(0)}(x^k)$ is an arbitrary function of the spatial coordinates
$\{x^k\}$ ($k=1,2,3$) and 
$f^{(0)}_{ij}(x^k)$ is a ($3\times 3$)-matrix function of the 
spatial coordinates with a unit determinant, respectively. 
Throughout 
this paper, this leading-order of spatial metric can be chosen as 
\begin{eqnarray}
f^{(0)}_{ij}=\delta_{ij},
\label{eq: f0-ij}
\end{eqnarray}
consistent with the gauge condition of (\ref{eq: gamma-ij t-infty}). On the other hand, $\ell^{(0)}$ represents a conserved comoving curvature perturbation, 
equivalent to 
a fluctuation of the number of $e$-folds, which is denoted by the so-called 
$\delta N$ term from some 
final uniform density (or comoving) hypersurface to the
initial flat hypersurface at $t=t_*$,
\begin{eqnarray}
\ell^{(0)}=\delta N(t_*,x^i)\,.
\label{eq: delta-N-u0}
\end{eqnarray} 

With these definitions of $H_LY$ and $H_TY$, we can define the nonlinear 
curvature perturbation valid up through $O(\epsilon^2)$ as
\begin{eqnarray}
{\cal R}^{\rm NL}_c\,\equiv\, {\zeta}_c\,+\,{E_c\over 3}\,.
\label{def0: nonlinear variable zeta}
\end{eqnarray}
It is easy to show that this nonlinear quantity can be reduced to 
(\ref{def: linear curvature}) in the linear limit. 
As clear from (\ref{def: nonlinear HT0}), finding $H_TY$ generally
requires a spatially non-local operation, however, 
in the comoving slicing, time-orthogonal gauge
with the asymptotic condition on the spatial 
coordinates~(\ref{eq: gamma-ij t-infty}), we find it
is possible to obtain the explicit form of $H_TY$ 
without any non-local operation as seen in \cite{Takamizu:2010xy}. 

Next, we can derive 
a nonlinear second-order differential equation that 
${\cal R}_c^{\rm NL}$ (\ref{def0: nonlinear variable zeta}) satisfies at 
$O(\epsilon^2)$ accuracy by introducing the conformal time $\eta$, defined by 
$d\eta={dt/a(t)}$ and 
the Mukhanov-Sasaki variable \cite{Mukhanov:1990me},
\begin{eqnarray}
z={{a\over H}\left(\rho+P \over c_s^2\right)^{1\over 2}},
\label{def: variable-z}
\end{eqnarray}
where notice that $c_s$ is the speed of sound for the gauge invariant scalar
perturbation in the linear theory~\cite{Garriga:1999vw}, given by 
%
\begin{equation}
 c_s^2 = \frac{P_X}{P_X+2P_{XX}X},
\end{equation}
where the subscript $X$ represents derivative with respect to $X$. 
The result can be reduced to a simple equation of the form 
(\ref{eq: basic eq for NL}) 
as a natural extension of the linear version 
(\ref{eq: linear eq R}). We also obtain 
the solution of the nonlinear equation (\ref{eq: basic eq for NL}) as
\begin{eqnarray}
{\cal R}_c^{\rm NL}(\eta)=&&\ell^{(0)}+
{1\over 4}\bigl[F(\eta)-F_*\bigr] 
K^{(2)}\nonumber\\ &&+ \bigl[D(\eta)-D_*\bigr] C^{(2)}+O(\epsilon^4), 
\label{eq: sol-tild-zeta}
\end{eqnarray}
where 
\begin{eqnarray}
&&D(\eta)=3{\cal H}_* \int_{\eta}^{0}{z^2(\eta_*)\over z^2(\eta')}d\eta'\,, \nonumber\\
&&F(\eta)=\int_{\eta}^{0} 
{d\eta'\over z^2(\eta')}
\int_{\eta_*}^{\eta'} z^2 c_s^2 (\eta'') d\eta''\,.
\label{def: integral-D-F}
\end{eqnarray}
Here $D_*=D(\eta_*)$, $F_*=F(\eta_*)$
and ${\cal H}_*$ denotes the conformal Hubble parameter
 ${\cal H}=d\ln a/d\eta$ at $\eta=\eta_*$ which 
 we take the time as some after the horizon crossing. 
Note that $t\to\infty$ corresponds to $\eta\to0$ in the conformal time.
Thus the functions $D$ and $F$ vanish asymptotically at late times,
$D(0)=F(0)=0$. 
Deviation of the solution (\ref{eq: sol-tild-zeta}) 
can be easily understood as follows. 
The second-order 
differential equation (\ref{eq: basic eq for NL}) 
contains two solutions (even though its independent relation 
appears only for the linear theory), i.e. decaying mode and growing mode. 
We can find that the function $D(\eta)$ satisfies 
\begin{eqnarray}
D''+2\frac{z'}{z}D'=0\,,
\label{eq: second-diff-D}
\end{eqnarray}
in the long-wavelength limit, 
i.e. no source term in (\ref{eq: basic eq for NL}). It will be seen 
that it corresponds to the decaying mode 
in the linear theory in the next section. On the 
other hand, the function $F(\eta)$ 
corresponds to the source term in (\ref{eq: basic eq for NL}), 
satisfying 
\begin{eqnarray}
F''+2\frac{z'}{z}F'+c_s^2=0\,,
\label{eq: second-diff-F}
\end{eqnarray}
as the $O(\epsilon^2)$ 
correction to a constant mode at the leading-order, i.e. as the 
growing mode in the linear theory, which is taken the 
form $1+F(\eta) K^{(2)}+O(\epsilon^4)$. 

Moreover the equation (\ref{eq: basic eq for NL}) includes 
two 'constants' of integration, or arbitrary spatial 
functions, which in general appear as the initial conditions, 
namely, the initial value and its time derivative. Let us consider the spatial 
functions, which we have introduced as 
$\ell^{(0)}, C^{(2)}$ and 
$K^{(2)}$. Here the last one is 
related to the Ricci scalar of the $0$th-order spatial metric as 
\begin{align}
K^{(2)}[\ell^{(0)}]=\, &R\left[e^{2 \ell^{(0)}}\delta_{ij}\right]\nonumber\\
=\, &
{-2 (2 \Delta \ell^{(0)}+\delta^{ij}\partial_i \ell^{(0)}
\partial_j \ell^{(0)})e^{-2 \ell^{(0)}}},
\label{def: K2}
\end{align}
where we have used $f^{(0)}_{ij}=\delta_{ij}$ from (\ref{eq: f0-ij}). Then we 
have the two arbitrary spatial functions: 
$\ell^{(0)}$ and $C^{(2)}$, which are related to 
the number of physical degrees of freedom for the initial conditions. 
Therefore $\ell^{(0)}$ and $C^{(2)}$ correspond to 
the initial conditions determined by matching a solution of 
$n$-th order perturbation solved inside the horizon to this 
superhorizon solution at $\eta=\eta_*$. 
Notice that $\ell^{(0)}$ represents $\delta N$ term as seen in 
(\ref{eq: delta-N-u0}) and $C^{(2)}$ originally comes from the decaying mode 
of the fluctuation of the scalar field \cite{Takamizu:2010xy}. 
\section{Growing and Decaying modes}
In this section, firstly, 
let us consider the growing and decaying modes in the linear theory. 
In the linear theory, 
the curvature perturbation on comoving hypersurfaces
follows (\ref{eq: linear eq R}). As usual, we consider it in 
Fourier space,
\begin{eqnarray}
{{\cal R}^{\rm Lin}_c}''+2{z'\over z} {{\cal R}^{\rm Lin}_c}'
+c_s^2k^2\,{\cal R}^{\rm Lin}_c=0\,.
\label{eq: linearFeq R}
\end{eqnarray}
Real space expressions (\ref{eq: linear eq R}) 
will be recovered by the replacement $k^2\to-\Delta$. 
This equation has two independent solutions, conventionally 
called a growing mode and a decaying mode. 

The growing mode
is a constant at the leading-order in the long-wavelength approximation 
or equivalently the spatial gradient expansion. 
Then in terms of the growing mode solution $u$, the 
decaying mode solution $v$ can be given as \cite{Leach:2001zf}
\begin{eqnarray}
v(\eta)=u(\eta){\tilde{D}(\eta)\over \tilde{D}(\eta_*)},
\tilde{D}(\eta)=3 {\cal H}_* \int_\eta^{0}
d\eta'{z^2(\eta_*)u^2(\eta_*)\over z^2(\eta')u^2(\eta')}\,.
\label{def: v-linear}
\end{eqnarray}
Note that this expression is correct for any order in the gradient 
expansion in the linear theory. 

The general solution of a curvature perturbation is 
written in terms of their linear combinations as 
\begin{eqnarray}
{\cal R}^{\rm Lin}_c(\eta)
 =\alpha^{\rm Lin} u(\eta)+\beta^{\rm Lin} v(\eta)\,,
\label{eq: linear sol-R0}
\end{eqnarray}
where the coefficients $\alpha^{\rm Lin}$ and $\beta^{\rm Lin}$ 
may be assumed to satisfy $\alpha^{\rm Lin}+\beta^{\rm Lin}=1$
without loss of generality. Note that the assumption of the 
gradient expansion (\ref{eqn:assumption-gamma}) corresponds to
the condition,
\begin{eqnarray}
\beta^{\rm Lin}=1-\alpha^{\rm Lin}=O(\epsilon^2)\,.
\label{eq: assumption-alpha}
\end{eqnarray}
This means, as mentioned before, that the decaying mode at
leading-order in the gradient expansion has already decayed
after horizon crossing.

Therefore the decaying mode solutions 
can be automatically obtained as following (\ref{def: v-linear}), 
if we obtain the growing mode solutions. 
Let us solve for the growing mode solution. 
In accordance with the gradient expansion, we set
\begin{eqnarray}
u(\eta)=\sum^\infty_{n=0} u_n(\eta) k^{2 n}\,.
\end{eqnarray}
At the leading-order in the gradient expansion, 
the growing mode solution $u^{(0)}$ is just a constant.
Then inserting the above expansion with $u^{(0)}=$const.
to the equation of motion~(\ref{eq: linearFeq R}) gives 
iteratively
\begin{eqnarray}
u''_{n+1}+2{z'\over z} u'_{n+1}=-c_s^2u_{n}\,.
\end{eqnarray}
As shown in \cite{Leach:2001zf}, 
$O(k^2)$ corrections to $u^{(0)}$ 
can be written as 
\begin{eqnarray}
u^{(2)}=u^{(0)}\left[C_1^{(2)} +C_2^{(2)} 
D(\eta)+k^2F(\eta)
\right],
\end{eqnarray}
where the integrals $D(\eta)$ and $F(\eta)$ have been given 
in (\ref{def: integral-D-F}), satisfying (\ref{eq: second-diff-D})
and (\ref{eq: second-diff-F}), respectively, as similar to the nonlinear 
theory, and  $C_1^{(2)}$ and $C_2^{(2)}$
are constants of integration. 
We fix the two constants as
$C_1^{(2)}=C_2^{(2)}=0$ so that $u^{(2)}$ is proportional to 
the integral $F(\eta)$ at $O(k^2)$ accuracy
\footnote{
If we fix the two arbitrary constants as
$C_1^{(2)}=0$ and $C_2^{(2)}=-k^2F_*/D_*$ so that
$u(\eta_*)=u^{(0)}$ holds at $O(k^2)$ accuracy, it is the case of \cite{Leach:2001zf} in which they discussed an enhancement of curvature perturbation on superhorizon scales due to suddenly change of the inflaton potential's 
slope, and its nonlinear effect also can be studied by matching 
the linear solution of \cite {Leach:2001zf} 
to our nonlinear solution in \cite{Takamizu:2010xy}}.
Hence we find 
\begin{eqnarray}
u^{(2)}(\eta)
=k^2u^{(0)}F(\eta)\,.
\label{sol: next-leading-grwoing}
\end{eqnarray}

As for the decaying mode, because of (\ref{eq: assumption-alpha}) 
we only need the leading-order solution. 
Since we may replace $\tilde D$ with $D$ in (\ref{def: v-linear}),
we immediately find 
\begin{eqnarray}
v^{(0)}=u^{(0)}{D(\eta)\over D_*}\,.
\label{eq: linear sol-R}
\end{eqnarray}
Thus from (\ref{sol: next-leading-grwoing}) and (\ref{eq: linear sol-R}), 
the general linear solution valid up through $O(\epsilon^2)$ is obtained 
as linear combination of constant $u^{(0)}$, growing mode $u^{(2)}$ and 
decaying mode $v^{(0)}$, which are proportional to $F(\eta)$ and $D(\eta)$, 
respectively. 

As for the nonlinear theory of cosmological perturbations, 
the solution of (\ref{eq: sol-tild-zeta}) 
is also shown as growing and decaying modes, respectively. 
We can find that the function $D(\eta)$ and $F(\eta)$ satisfy 
(\ref{eq: second-diff-D}) and (\ref{eq: second-diff-F}), respectively and 
they take the same forms, respectively as in the linear theory. Therefore we 
can interpret that they correspond to 
the decaying mode in the long-wavelength limit, 
and the growing mode taken 
the form $1+k^2F(\eta)+O(k^4)$, where $F(\eta)$ is 
the $k^2$ correction to the growing (i.e., constant) mode, respectively. 
In our nonlinear theory, note that 
time derivative takes the same form as shown in (\ref{eq: basic eq for NL}) and (\ref{eq: linear eq R}), hence the decaying mode can not couple 
with the growing mode as similar to the linear theory because of 
the method of gradient expansion (i.e. time derivative takes 
as a linear operator). 
The difference from the linear theory is the source term 
, i.e. the Ricci scalar of the leading order metric $K^{(2)}$, which can be 
reduced to $k^2 {\cal R}^{\rm Lin}_c$ in Fourier space as the source term in the linear theory 
(\ref{eq: linearFeq R}).
\section{Crossing of $\dot \phi=0$}
We consider the case when $\dot{\phi}$ (or $z$) crosses zero in our 
nonlinear theory. The gradient expansion allows us to discuss 
in a similar way as the linear theory \cite{Seto:1999jc,Leach:2001zf}. 
For simplicity, we assume that $z$ changes the sign only once at 
$\eta=\eta_0$. Hereafter, we consider a canonical single scalar field, however, the same discussion can be done in the case of 
a non-canonical single scalar field, when $P_X X \approx 0$. 

In the vicinity of $\eta=\eta_0$, $z$ can be expressed as 
\begin{eqnarray}
z=z_0'(\eta-\eta_0),
\end{eqnarray}
where $z_0'=z'(\eta_0)$. Hence the equation for ${\cal R}^{\rm NL}_c$ becomes
\begin{eqnarray}
\left[{d^2\over d \eta^2}+{2\over \eta-\eta_0} {d\over d\eta}\right]{\cal R}_c^{\rm NL}=\nonumber\\-{1\over 4} K^{(2)}[\,
{\cal R}_c^{\rm NL}\,]
+O(\epsilon^4).
\end{eqnarray}
The two independent solutions 
in the linear theory (not guaranteed in the nonlinear theory) 
can be found as 
\begin{align}
u\approx\,& \ell^{(0)}\left(1-{1\over 6}K^{(2)}[\,
{\cal R}_c^{\rm NL}\,](\eta-\eta_0)^2+\cdots\right),
\label{eq: growing}\\
v\approx\,& C^{(2)}\left({1\over \eta-\eta_0}-{1\over 2}K^{(2)}[\,
{\cal R}_c^{\rm NL}\,](\eta-\eta_0)+\cdots\right).
\label{eq: decaying}
\end{align}

We consider $u$ and $v$ should be chosen as the growing mode and 
decaying mode, respectively, and $u$ remains constant across the epoch $\eta=\eta_0$. The second term in (\ref{eq: growing}) can be obtained by 
the integral $F(\eta)$, which in this case is given by 
\begin{eqnarray}
F(\eta)\propto \lim_{\eta\to \eta_0} (\eta-\eta_0)^2,
\end{eqnarray}
and shown to be still well defined in the crossing of $\dot{\phi}=0$.  
The final value of the growing (or non-decaying) mode 
at late times will take a constant $\ell^{(0)}$ (i.e. $\delta N$ term). 

The singularity will appear in the first term in (\ref{eq: decaying}), arising from the integral $D(\eta)$. It can be expressed as in 
the case of linear theory and for $\eta>\eta_0$, we obtain as 
\begin{eqnarray}
D(\eta)\propto \int^0_\eta {d \eta'\over z^2}\approx 
{1\over {z_0'}^2 (\eta-\eta_0)}.
\end{eqnarray}
We conclude that this term  diverges  
in the limit $\eta\to \eta_0+0$, however, this is just a decaying mode, 
then it will vanish definitely at late times as 
\begin{eqnarray}
D(\eta)\propto a^{-3} \to 0,~~~{\rm with}~~\eta\to 0.
\end{eqnarray}
Hence we can see that 
no problem will occur for both the linear and nonlinear theory. 
\section{Concluding remarks }
We clarify what happens when the inflaton stops during inflation for 
nonlinear cosmological perturbation theory. 
We focus on the evolution on 
the superhorizon scales and review our nonlinear theory, called the 
{\it beyond} $\delta N$-formalism for a general single scalar field as 
the next-leading order in the gradient expansion. 
In our nonlinear theory, we can deal with the 
time evolution in contrast to $\delta N$-formalism where 
curvature perturbations remain just constant. 

As a summary of 
our formula, 
note that time derivative takes the same form as shown in (\ref{eq: basic eq for NL}) and (\ref{eq: linear eq R}), hence the decaying mode can not couple 
with the growing mode as similar to the linear theory because of 
the method of gradient expansion, i.e. time derivative takes 
as a linear operator. 
The difference from the linear theory is the source term 
, i.e. the Ricci scalar of the leading-order metric $K^{(2)}$, which can be 
reduced to $k^2 {\cal R}^{\rm Lin}_c$ in Fourier space as the source term in the linear theory 
(\ref{eq: linearFeq R}). 

We can conclude that although the decaying mode diverges  
in the limit of time when $\dot{\phi}$ vanishes, there appears no trouble for 
both the linear and nonlinear theory since this mode will 
vanish definitely at late times. 
\section{acknowledgements}
This work was supported by JSPS Grant-in-Aid for Young Scientists (B) No. 21740192. and for Scientific Research No. 19340054.


\begin{thebibliography}{99}

\bibitem{Komatsu:2010fb}
  E.~Komatsu {\it et al.},
  arXiv:1001.4538 [astro-ph.CO].
\bibitem{Guth:1980zm}
  A.~H.~Guth,
  Phys.\ Rev.\  D {\bf 23} 347 (1981).

\bibitem{Starobinsky:1980te}
  A.~A.~Starobinsky,
  Phys.\ Lett.\  B {\bf 91} 99 (1980).

\bibitem{Sato:1980yn}
  K.~Sato,
  Mon.\ Not.\ Roy.\ Astron.\ Soc.\  {\bf 195} 467 (1981).

\bibitem{Mukhanov:1981xt}
  V.~F.~Mukhanov and G.~V.~Chibisov,
  JETP Lett.\  {\bf 33} 532 (1981) 
  [Pisma Zh.\ Eksp.\ Teor.\ Fiz.\  {\bf 33} 549 (1981)].

\bibitem{Hawking:1982cz}
  S.~W.~Hawking,
  Phys.\ Lett.\  B {\bf 115} 295 (1982).


\bibitem{Starobinsky:1982ee}
  A.~A.~Starobinsky,
  Phys.\ Lett.\  B {\bf 117} 175 (1982).

\bibitem{Guth:1982ec}
  A.~H.~Guth and S.~Y.~Pi,
  Phys.\ Rev.\ Lett.\  {\bf 49} 1110 (1982).

\bibitem{Damour:1997cb}
  T.~Damour and V.~F.~Mukhanov,
  Phys.\ Rev.\ Lett.\  {\bf 80}, 3440 (1998).

\bibitem{Liddle:1998pz}
  A.~R.~Liddle and A.~Mazumdar,
  Phys.\ Rev.\  D {\bf 58}, 083508 (1998); 
  A.~Taruya,
  Phys.\ Rev.\  D {\bf 59}, 103505 (1999).


\bibitem{Yokoyama:1998pt}
  J.~Yokoyama,
  Phys.\ Rev.\  D {\bf 58}, 083510 (1998);
{\bf 59}, 107303 (1999).


\bibitem{Seto:1999jc}
  O.~Seto, J.~Yokoyama and H.~Kodama,
  Phys.\ Rev.\  D {\bf 61}, 103504 (2000).

\bibitem{Kodama:1996jh}
  H.~Kodama and T.~Hamazaki,
  Prog.\ Theor.\ Phys.\  {\bf 96}, 949 (1996).

\bibitem{Salopek:1990jq}
  D.~S.~Salopek and J.~R.~Bond,
  Phys.\ Rev.\  D {\bf 42}, 3936 (1990).


\bibitem{Starobinsky:1986fxa}
  A.~A.~Starobinsky,
  JETP Lett.\  {\bf 42}, 152 (1985)
  [Pisma Zh.\ Eksp.\ Teor.\ Fiz.\  {\bf 42}, 124 (1985)].

\bibitem{Nambu:1994hu}
  Y.~Nambu and A.~Taruya,
  Class.\ Quant.\ Grav.\  {\bf 13}, 705 (1996).


\bibitem{Sasaki:1995aw}
  M.~Sasaki and E.~D.~Stewart,
  Prog.\ Theor.\ Phys.\  {\bf 95}, 71 (1996).

\bibitem{Lyth:2005fi}
  D.~H.~Lyth and Y.~Rodriguez,
  Phys.\ Rev.\ Lett.\  {\bf 95}, 121302 (2005).

\bibitem{Bartolo:2004if}
  N.~Bartolo, E.~Komatsu, S.~Matarrese and A.~Riotto,
  Phys.\ Rept.\  {\bf 402}, 103 (2004). 

\bibitem{Takamizu:2010xy}
  Y.~Takamizu, S.~Mukohyama, M.~Sasaki and Y.~Tanaka,
  JCAP {\bf 1006}, 019 (2010).

\bibitem{Takamizu:2008ra}
  Y.~Takamizu and S.~Mukohyama,
  JCAP {\bf 0901}, 013 (2009).

\bibitem{Mukhanov:1990me}
  V.~F.~Mukhanov, H.~A.~Feldman and R.~H.~Brandenberger,
  Phys.\ Rept.\  {\bf 215}, 203 (1992). 

\bibitem{Kodama:1985bj}
  H.~Kodama and M.~Sasaki,
  Prog.\ Theor.\ Phys.\ Suppl.\  {\bf 78} 1 (1984).

\bibitem{Garriga:1999vw}
  J.~Garriga and V.~F.~Mukhanov,
  Phys.\ Lett.\  B {\bf 458}, 219 (1999).

\bibitem{Leach:2001zf}
  S.~M.~Leach, M.~Sasaki, D.~Wands and A.~R.~Liddle,
  Phys.\ Rev.\  D {\bf 64}, 023512 (2001).
\end{thebibliography}
\end{document}